\documentclass[aps, prd, twocolumn, lengthcheck, superscriptaddress, showpacs, letterpaper, nofootinbib]{revtex4-1}

\usepackage{amsmath,amssymb}
\usepackage{graphicx,bm}
\usepackage{epstopdf}
\usepackage{ulem}
\usepackage[usenames]{color}
\usepackage{float}
\usepackage{multirow}
\allowdisplaybreaks

\def\vs{{v\hspace{-0.17cm}\slash}}

\begin{document}

\title{Chiral partner structure of doubly heavy baryons with heavy quark spin-flavor symmetry }

\author{Yong-Liang Ma}
\email{yongliangma@jlu.edu.cn}
\affiliation{Center for Theoretical Physics and College of Physics, Jilin University, Changchun, 130012, China}

\author{Masayasu Harada}
\email{harada@hken.phys.nagoya-u.ac.jp}
\affiliation{Department of Physics, Nagoya University, Nagoya, 464-8602, Japan}

\date{\today}

\begin{abstract}
The spectrum of the doubly heavy baryons is estimated with respect to the new observation from the LHCb collaboration~\cite{Aaij:2017ueg} by using the chiral partner structure and heavy quark spin-flavor symmetry. The effects of heavy quark flavor symmetry breaking and light quark flavor symmetry are considered. The mass splitting of doubly heavy baryons with the same spin but opposite parity arises from the generalized Goldberger-Treiman relation. The intermultiplet one-pion transition and intramultiplet radiative transitions are also be estimated.
\end{abstract}
% 11.30.Hv	Flavor symmetries
% 11.30.Rd Chiral symmetries
% 12.39.Hg	Heavy quark effective theory
% 14.20.Lq	Charmed baryons (|C|>0, B=0)
% 14.20.Mr	Bottom baryons (|B|>0)

\maketitle

%%%%%%%%%%%%%%%%%%%%%%%%%%%%%

Very recently, the doubly charmed baryon $\Xi_{cc}^{++}$ was observed by the LHCb collaboration~\cite{Aaij:2017ueg} at a very high confidence level. Differently from the experimental status of its light flavor partner of $\Xi_{cc}^{+}$, the observation of $\Xi_{cc}^{++}$ confirms the existence of the doubly heavy baryons (DHBs) which are the straight forward prediction from the extension of the quark model~\cite{GellMann:1964nj,Zweig:1981pd,Zweig:1964jf,DeRujula:1975qlm}. It is reasonable to believe that this mile stone observation unveils a new direction in particle physics. After the observation of LHCb collaboration, properties of DHBs have been considered by several groups~\cite{Chen:2017sbg,Li:2017cfz,Wang:2017mqp}.

Inside a DHB, there is a heavy diquark and a light quark. Since the heavy quark in a DHB is almost near its mass shell, it is reasonable to expect that the heavy quark limit is applicable in this system (for a review, see, e.g., \cite{Manohar:2000dt}). Since there is a light quark in a DHB, the chiral symmetry which is widely used in the physics of light hadrons is applicable. In this paper, concerning the experimental progress~\cite{Aaij:2017ueg} and our previous works~\cite{Ma:2015lba,Ma:2015cfa}, we devote ourselves to study the spectrum of the DHBs by using the effective model with respect to the chiral symmetry and heavy quark spin-flavor symmetry.

Since the heavy quark has a large mass, in the heavy quark limit, the heavy diquark in a DHB can be regarded as a compact object without radical excitation, that is, either a boson $\Phi$ with quantum numbers $J_Q^P = 0^+$ or a boson $\Phi_\mu$ with quantum numbers $1^+$~\footnote{Note that, for a heavy diquark composed by the same heavy quarks, $\Phi$ does not exist. In addition, in the present work, we changed our convention $\bar{\Phi}_{(\mu)}$ used in the previous works ~\cite{Ma:2015lba,Ma:2015cfa} to $\Phi_{(\mu)}$.}. Schematically, the quark contents of these two states are
\begin{eqnarray}
\Phi  \sim  Q^T \gamma_5 C Q^\prime\ , \quad
\Phi_\mu \sim Q^T \gamma_\mu C Q^\prime\ ,
\end{eqnarray}
where $\Phi$ and $\Phi_\mu$ include the annihilation operator for a heavy diquark, $C = i\gamma_2\gamma_0$ is the charge conjugation operator and the heavy quark flavor matrix $Q^{(\prime)} = (c,\, b)$. Both states are color anti-triplets. And, in the heavy quark flavor basis, we have the following matrix form
\begin{eqnarray}
\Phi & = & \begin{pmatrix} 0 & \Phi_{cb} \\ \Phi_{bc} & 0 \\ \end{pmatrix} \ ,
\quad
\Phi^{\mu} = \begin{pmatrix} \Phi_{cc}^{\mu} & \Phi_{bc}^{\mu} \\ \Phi_{bc}^{\mu} & \Phi_{bb}^{\mu} \\
 \end{pmatrix} \ ,
\end{eqnarray}
in which $\Phi_{cb} ={} - \Phi_{bc}$ due to the heavy quark flavor asymmetry. Because of the spin-flavor symmetry in heavy quark limit~\cite{Isgur:1989vq,Isgur:1989ed,Shifman:1986sm,Shifman:1986sm,Shifman:1987rj}, all the states in $\Phi$ and $\Phi_\mu$ have the same mass.

In the heavy quark limit, these states can be arranged into a heavy quark multiplet
\begin{equation}
X =  P_+\left( \gamma_5 C \Phi + i\gamma^\mu C \Phi_\mu\right)  P_+^T,
\end{equation}
where $P_{\pm} = (1 \pm \vs)/2$. Since colors of two heavy quarks in $X$ are antisymmetric, the spin and flavor indices are taken to be symmetric. Then the representation of $X$ under the heavy quark spin-flavor SU(4) symmetry is $\boldmath 10$, which is decomposed into $({\boldmath 1},{\boldmath 1}) \oplus ({\boldmath 3},{\boldmath 3})$ corresponding to $\Phi \,\oplus \Phi_\mu$.
The parity transformation of $X$ is written as
\begin{equation}
X \ \mathop{\to}_P \ \gamma_0 X \gamma_0 \ .
\end{equation}
After an appropriate normalization, the anti-heavy quark field $\bar{Q}$ and the heavy diquark field $X$ can be arranged into a column
\begin{eqnarray}
\left(
                         \begin{array}{c}
                           \bar{Q} \\
                           X \\
                         \end{array}
                       \right)
.
\end{eqnarray}
And the color interaction does not mix $\bar{Q}$ and $X$. That is, there is a superflavor symmetry for the heavy quark sector~\cite{Georgi:1990ak,Carone:1990pv}.

We next couple the light quark to the heavy object  to study the heavy hadrons including a light quark. With respect to the light quark components of heavy hadrons, in addition to the heavy quark spin-flavor symmetry, the chiral partner structure can be used to classify the heavy hadrons, like what was proposed for heavy-light mesons~\cite{Nowak:1992um,Bardeen:1993ae,Bardeen:2003kt,Nowak:1993vc}. For studying the chiral partner structure of the heavy hadrons, we consider light-quark clouds with $j^P = (1/2)^+$ and $(1/2)^-$ as parity partners to each other where the former comes from the $s$-wave coupling between the light quark and heavy constituent but the latter comes from the $p$-wave coupling.

Considering these two light-quark clouds with $j^P = (1/2)^+$ and $(1/2)^-$, one can define two kinds of heavy hadrons, one kind $\Psi_{+}$ with a light-quark cloud $j^P = (1/2)^+$  and the other kind $\Psi_{-}$ with a light-quark cloud $j^P = (1/2)^-$ as follows
\begin{eqnarray}
\Psi_+ & = & \left(
                         \begin{array}{c}
                           \bar{H} \\
                           \mathcal{B}_+ \\
                         \end{array}
                       \right), \quad \Psi_- = \left(
                         \begin{array}{c}
                           \bar{G} \\
                           \mathcal{B}_- \\
                         \end{array}
                       \right),
\end{eqnarray}
where $\bar{H}$ and $\bar{G}$ include anti-heavy mesons  with $J^P = (0^-,1^-)$ and $(0^+,1^+)$,
respectively, and the fields ${\mathcal B}_{+}$ and ${\mathcal B}_{-}$ include DHBs with positive and negative parities, respectively.

By writing the anti-heavy meson fields with $J^P = (0^-,1^-)$ and $(0^+,1^+)$
as $(\bar{P}, \bar{P}^{\ast \mu})$ and $(\bar{P}_0^\ast,\bar{P}_1^\mu)$, $\bar{H}$ and $\bar{G}$ are expressed as
\begin{eqnarray}
\bar{H} & = & \left( \bar{P}^{\ast\mu}\gamma_\mu +  i \bar{P} \gamma_5 \right) P_+
, \nonumber\\
\bar{G} & = & \left({} - \bar{P}_1^{\mu} \gamma_\mu \gamma_5 + \bar{P}_0^\ast \right) P_+
 \ .
\label{eq:HGphys}
\end{eqnarray}
It should be noted that $\bar{P}$ is doublet of the heavy flavor symmetry including $\bar{D}$ and $B$ mesons and triplet under the light flavor symmetry:

\begin{eqnarray}
\bar{P} = \begin{pmatrix} \bar{D}^0  & B^+ \\  D^- & B^0 \\ D_s^- & B_s^0 \end{pmatrix} \ ,
\end{eqnarray}
and similarly for $\bar{P}^{\ast \mu}$, $\bar{P}_0^\ast$ and $\bar{P}_1^\mu$.

Because of the existence of the heavy flavor symmetry, ${\mathcal B}_{\pm}$ includes a heavy-quark spin singlet with $J^P = \frac{1}{2}^\pm$ and a heavy-quark spin doublet with $J^P = \left( \frac{1}{2}^\pm , \frac{3}{2}^\pm \right)$. By writing the singlet field as $\phi_{\pm}$ and the doublet field as $\psi^\mu_{\pm}$, ${\mathcal B}_{\pm}$ are expressed as
\begin{eqnarray}
& & \left[ \mathcal{B}_{+} \right]_{h h' l} = \left[ P_+ \right]_{h h_1} \Big( \left[ \gamma_5 C  \right]_{h_1 h_2} \left[ \phi_{+} \right]_l \nonumber\\
& &\qquad\qquad\qquad\qquad\quad{} + \left[ \gamma_\mu C \right]_{h_1 h_2}  \left[ \psi_{+}^{\mu} \right]_l \Big) \left[ P_+^T \right]_{h_2 h'} \ , \\
& & \left[ \mathcal{B}_{-} \right]_{h h' l} = \left[ P_+ \right]_{h h_1} \Big( \left[ \gamma_5 C  \right]_{h_1 h_2} \left[ \gamma_5 \phi_{-} \right]_l \nonumber\\
& &\qquad\qquad\qquad\qquad\quad{} + \left[ \gamma_\mu C \right]_{h_1 h_2}  \left[ \gamma_5 \psi_{-}^{\mu} \right]_l \Big) \left[ P_+^T \right]_{h_2 h'} \ ,
\nonumber
\end{eqnarray}
where $h$, $h'$, $h_1$, $h_2$ are spinor indices for heavy quarks and $l$ is the spinor index for baryons.  The summation over repeated indices are understood.
For simplicity of writing, we express the above structure as
\begin{align}
{\mathcal B}_{+} & = P_+ \left[ \gamma_5 C \otimes \phi_+ + \gamma_\mu C \otimes \psi_+^\mu \right] P_+^T   \ , \notag\\
{\mathcal B}_{-} & = P_+ \left[ \gamma_5 C \otimes \gamma_5 \phi_- + \gamma_\mu C \otimes \gamma_5 \psi_-^\mu \right] P_+^T   \ .
\end{align}

We further decompose the $\psi_{\pm}^\mu$ fields into $J^P = \frac{1}{2}^\pm$ field $\psi_{\pm\,1/2}$ and $J^P = \frac{3}{2}^\pm$ field $\psi_{\pm \, 3/2}^\mu$ through~\cite{Carone:1990pv}
\begin{eqnarray}
\psi_{\pm}^\mu = \frac{1}{\sqrt{3}} \sigma^{\mu\nu} \, v_\nu \psi_{\pm \, 1/2} + \psi_{\pm \, 3/2}^\mu \ ,
\end{eqnarray}
where
\begin{equation}
\sigma^{\mu\nu} = \frac{i}{2} \left[ \gamma^\mu \,,\, \gamma^\nu \right] \ ,
\end{equation}
and the spin-$3/2$ Rarita-Schwinger fields $\psi_{\pm\,3/2}^\mu$ satisfy
\begin{equation}
\vs \psi_{\pm\,3/2}^\mu = \psi_{\pm \, 3/2}^\mu \ , \quad v_\mu \psi_{\pm\,3/2}^\mu = \gamma_\mu \psi_{\pm\,3/2}^\mu = 0 \ .
\end{equation}
Each field of ${\mathcal B}_{\pm}$ carries three flavor indices, two of which are for heavy flavors, $c$ and $b$, and one for light flavors, $u$, $d$, $s$.
For clarifying notations, we express $\phi_{\pm}$, $\psi_{\pm\,1/2}$ and $\psi_{\pm\,3/2}^\mu$ in terms of physical baryons:
\begin{eqnarray}
\phi_+ & = & \begin{pmatrix} 0 & - \left( \Xi_{bc}^{\prime+} , \Xi_{bc}^{\prime0} , \Omega_{bc}^{\prime0} \right) \\
\left( \Xi_{bc}^{\prime+} , \Xi_{bc}^{\prime0} , \Omega_{bc}^{\prime0} \right) & 0 \\ \end{pmatrix} \ , \nonumber\\
\phi_- & = & \begin{pmatrix} 0 & - \left( \Xi_{bc}^{\prime\ast +} , \Xi_{bc}^{\prime\ast 0} , \Omega_{bc}^{\prime\ast 0} \right) \\
\left( \Xi_{bc}^{\prime\ast +} , \Xi_{bc}^{\prime\ast 0} , \Omega_{bc}^{\prime\ast 0} \right) & 0 \\ \end{pmatrix} \ , \nonumber\\
\psi_{+\,1/2} & = & \begin{pmatrix} \left( \Xi_{cc}^{++} , \Xi_{cc}^{+} , \Omega_{cc}^{+} \right)  &  \left( \Xi_{bc}^{+} , \Xi_{bc}^{0} , \Omega_{bc}^{0} \right) \\
\left( \Xi_{bc}^{+} , \Xi_{bc}^{0} , \Omega_{bc}^{0} \right) & \left( \Xi_{bb}^0 , \Xi_{bb}^- , \Omega_{bb}^- \right) \\ \end{pmatrix} \ , \nonumber\\
\psi_{-\,1/2} & = & \begin{pmatrix} \left( \Xi_{cc}^{\ast++} , \Xi_{cc}^{\ast+} , \Omega_{cc}^{\ast+} \right)  &  \left( \Xi_{bc}^{\ast+} , \Xi_{bc}^{\ast0} , \Omega_{bc}^{\ast0} \right) \\
\left( \Xi_{bc}^{\ast+} , \Xi_{bc}^{\ast0} , \Omega_{bc}^{\ast0} \right) & \left( \Xi_{bb}^{\ast0} , \Xi_{bb}^{\ast-} , \Omega_{bb}^{\ast-} \right) \\ \end{pmatrix} \ , \\
\psi_{+\,3/2}^\mu & = & \begin{pmatrix} \left( \Xi_{cc}^{\mu++} , \Xi_{cc}^{\mu+} , \Omega_{cc}^{\mu+} \right)  &  \left( \Xi_{bc}^{\mu+} , \Xi_{bc}^{\mu0} , \Omega_{bc}^{\mu0} \right) \\
\left( \Xi_{bc}^{\mu+} , \Xi_{bc}^{\mu0} , \Omega_{bc}^{\mu0} \right) &  \left( \Xi_{bb}^{\mu0} , \Xi_{bb}^{\mu-} , \Omega_{bb}^{\mu-} \right) \\ \end{pmatrix} \ , \nonumber\\
\psi_{-\,3/2}^\mu & = & \begin{pmatrix} \left( \Xi_{cc}^{\prime\mu++} , \Xi_{cc}^{\prime\mu+} , \Omega_{cc}^{\prime\mu+} \right)  &  \left( \Xi_{bc}^{\prime\mu+} , \Xi_{bc}^{\prime\mu0} , \Omega_{bc}^{\prime\mu0} \right) \\
\left( \Xi_{bc}^{\prime\mu+} , \Xi_{bc}^{\prime\mu0} , \Omega_{bc}^{\prime\mu0} \right) &  \left( \Xi_{bb}^{\prime\mu0} , \Xi_{bb}^{\prime\mu-} , \Omega_{bb}^{\prime\mu-} \right) \\ \end{pmatrix} \ ,\nonumber
\end{eqnarray}

We note that the parity transformations of $\bar{H}$, $\bar{G}$ and ${\mathcal B}_\pm$ are given by
\begin{eqnarray}
& & \bar{H} \ \mathop{\to}_P \ \gamma_0 \bar{H} \gamma_0 \ , \quad  \bar{G} \ \mathop{\to}_P \ \gamma_0 \bar{G} \gamma_0 \ , \notag\\
& & \left[ \mathcal{B}_{\pm} \right]_{h h' l} \ \mathop{\to}_P \
\left[ \gamma_0 \right]_{l l_1}
\left[ \gamma_0 \right]_{h h_1} \left[ \mathcal{B}_{\pm} \right]_{h_1 h_2 l_2}  \left[ \gamma_0 \right]_{h_2 h'} \ .
\end{eqnarray}
For later convenience, we define the conjugation fields of $\Psi_{\pm}$ as
\begin{eqnarray}
\bar{\Psi}_+ = \begin{pmatrix}
	H \\ \bar{\mathcal B}_+
\end{pmatrix} \ , \quad
\bar{\Psi}_- = \begin{pmatrix}
	G \\ \bar{\mathcal B}_-
\end{pmatrix},
\end{eqnarray}
where
\begin{eqnarray}
& & H = \gamma_0 \bar{H}^\dag \gamma_0 \ , \quad G = \gamma_0 \bar{G}^\dag \gamma_0 \ , \nonumber\\
& & \left[ \bar{\mathcal B}_{\pm} \right]_{hh' l} = \left[ \gamma_0 \right]_{hh_1} \left[ {\mathcal B}_{\pm}^\dag \right]_{h_1h_2 l_1} \left[ \gamma_0 \right]_{h_2h'} \left[ \gamma_0 \right]_{l_1 l} .
\end{eqnarray}
By writing spinor structure explicitly, these fields are expressed as
\begin{eqnarray}
H & = & P_+ \left[ \left(\bar{P}^{\ast\mu}\right)^\dag\gamma_\mu +  i \left(\bar{P}\right)^\dag \gamma_5 \right] \
, \nonumber\\
G & = & P_+ \left[{} - \left(\bar{P}_1^{\prime\mu}\right)^\dag \gamma_\mu \gamma_5 + \left(\bar{P}_0^\ast\right)^\dag \right]
 \ ,
\end{eqnarray}
and
\begin{eqnarray}
\bar{\mathcal B}_{+} & = & P_+^T \left[ C \gamma_5 \otimes \bar{\phi}_+ + C \gamma_\mu \otimes \bar{\psi}_+^\mu \right] P_+   \ , \nonumber\\
\bar{\mathcal B}_{-} & = & P_+^T \left[{} - C \gamma_5 \otimes \bar{\phi}_- \gamma_5 - C \gamma_\mu \otimes \bar{\psi}_-^\mu \gamma_5 \right] P_+   \ .
\end{eqnarray}

In terms of the parity eigenstates, one can make the combination
\begin{eqnarray}
\Psi_{L} & = & \frac{1}{\sqrt{2}}\left(\Psi_{+} - i \gamma_5 \Psi_{-}\right)
\ , \nonumber\\
\Psi_{R} & = & \frac{1}{\sqrt{2}}\left(\Psi_{+} + i \gamma_5 \Psi_{-} \right)\ ,
\label{eq:ChiralPhys}
\end{eqnarray}
where $\gamma_5$ act on the index corresponding to light quarks. For later convenience, we introduce
\begin{eqnarray}
\bar{\mathcal H}_L & = & \frac{1}{\sqrt{2}} \left( \bar{H} - i \gamma_5 \bar{G} \right) \ , \quad
\bar{\mathcal H}_R = \frac{1}{\sqrt{2}} \left( \bar{H} + i \gamma_5 \bar{G} \right) \ , \nonumber\\
{\mathcal B}_L & = & \frac{1}{\sqrt{2}} \left( {\mathcal B}_+ - i \gamma_5 {\mathcal B}_- \right) , \;  {\mathcal B}_R = \frac{1}{\sqrt{2}} \left( {\mathcal B}_+ + i \gamma_5 {\mathcal B}_- \right) .
\nonumber\\
\end{eqnarray}
By using these fields, $\Psi_{L,R}$ are expressed as
\begin{equation}
\Psi_{L,R} = \left(
               \begin{array}{c}
                 {\mathcal H}_{L,R} \\
                  {\mathcal B}_{L,R}  \\
               \end{array}
             \right)\ .
\end{equation}
Under chiral transformation, $\Psi_{L,R}$ transform in the same way as the current chiral quark, i.e.,
\begin{eqnarray}
\Psi_{L,R} & \to & g_{L,R} \Psi_{L,R},
\end{eqnarray}
with $g_{L,R}\in SU(3)_{L,R}$.

%\begin{widetext}
Now, we are in the position to construct the chiral effective Lagrangian for heavy hadrons with chiral partner structure. Following the procedure used in Refs.~\cite{Ma:2015lba,Ma:2015cfa} the effective Lagrangian preserving chiral symmetry as well as heavy quark symmetry can be written as
\begin{widetext}
\begin{eqnarray}
{\cal L}_{\rm \Psi} & = &{} - {\rm Tr}\left(\bar{\Psi}_{L} i v\cdot \partial \Psi_{L} +
\bar{\Psi}_{R} i v\cdot \partial \Psi_{R}\right) + \Delta{\rm Tr}\left( \bar{\Psi}_{L} \Psi_{L} + \bar{\Psi}_{R}
\Psi_{R} \right) \nonumber\\
& &{} - \frac{1}{2}g_\pi{\rm Tr}\left( \bar{\Psi}_{L} M \Psi_{R} +
\bar{\Psi}_{R} M^\dagger \Psi_{L} \right) + i\frac{g_A}{f_\pi}{\rm Tr}\left( \bar{\Psi}_{L}\gamma_5\gamma^\nu \partial_\nu M
\Psi_{R}  - \bar{\Psi}_{R}\gamma_5\gamma^\nu \partial_\nu M^\dagger
\Psi_{L}\right)\nonumber\\
& &{} + i \frac{( g_S + i g_I ) }{4 f_\pi^2} \mbox{Tr} \left[ \bar{\Psi}_L \left( v\cdot \partial M \right) M^\dag \Psi_L  + \bar{\Psi}_R \left( v\cdot \partial M^\dag \right) M \Psi_R \right] \nonumber\\
& &{} -
 i \frac{( g_S - i g_I ) }{4 f_\pi^2} \mbox{Tr} \left[ \bar{\Psi}_L M \left( v\cdot \partial M^\dag \right) \Psi_L  + \bar{\Psi}_R M^\dag \left( v\cdot \partial M \right) \Psi_R \right] ,
\label{eq:EffecL}
\end{eqnarray}
\end{widetext}
where $M$ is the light meson field which transforms as $M \to g_L M g_R^\dag$ under chiral transformation,
and $g_\pi$, $g_A$, $g_S$ and $g_I$ are real dimensionless coupling constants.
%\end{widetext}
In terms of the scalar and pseudoscalar fields, one can make a decomposition $M = S + i \Phi = 2S^a T^a + 2 i \left( \pi^a T^a \right)$
with $S^a$ being the scalar nonet field, $\pi^a$ being the pseudoscalar nonet field and $T^a$ being the generators of U(3) group with the normalization $\mbox{tr} \left( T_a T_b \right) = (1/2) \delta^{ab}$. Compared to our previous works~\cite{Ma:2015lba,Ma:2015cfa}, we include the $g_S$ term in Eq.~\eqref{eq:EffecL} which is significant for predicting the intermultiplet hadronic decay.
%\end{widetext}

As we know that, even though the charm quark mass and bottom quark mass are both large compared to their off-shell scales, the flavor symmetry is strongly broken. To account for the mass splitting between charm baryon and bottom baryon with preserving the heavy quark spin symmetry, we introduce
\begin{eqnarray}
{\cal L}_{\Psi}^{\mathcal{M}_Q}  & = & - c_1\, \mbox{Tr} \left[ {\mathcal M}_Q^{-1} \, \left( {\mathcal H}_L \bar{\mathcal H}_L  +  {\mathcal H}_R \bar{\mathcal H}_R \right) \right]  \notag\\
& &{}
- c_2\, \mbox{Tr} \left[ \left( {\mathcal M}_Q^{-1} \right)^T \left( \bar{\mathcal B}_L {\mathcal B}_L  +  \bar{\mathcal B}_R {\mathcal B}_R \right) \right]  \notag\\
& &{}
- c_3\, \mbox{Tr} \left[ \bar{\mathcal B}_L \, {\mathcal M}_Q^{-1}  \, {\mathcal B}_L  +  \bar{\mathcal B}_R  \, {\mathcal M}_Q^{-1}  \, {\mathcal B}_R  \right]  \ ,
\label{eq:mQCor}
\end{eqnarray}
where $c_1$, $c_2$ and $c_3$ are constants with dimension mass square and $\mathcal{M}_Q^{-1} = {\rm diag}(1/m_c, 1/m_b)$ which recovers the flavor symmetry when $m_c \to m_b$.

Now, let us consider the mass spectra. With including the heavy flavor symmetry violation in Eq.~(\ref{eq:mQCor}),
the masses of the heavy-light mesons are expressed as
\begin{eqnarray}
M_{H_c} & = & \mathcal{M}_H - \frac{c_1}{m_c} \ , \quad M_{H_b} = \mathcal{M}_{H} - \frac
{c_1}{ m_b} \ ,\nonumber\\
M_{G_c} & = & \mathcal{M}_G +  \frac{c_1}{m_c} \ ,  \quad M_{G_b} = \mathcal{M}_{G} +  \frac
{c_1}{ m_b},
\end{eqnarray}
where $\mathcal{M}_H$ and $\mathcal{M}_G$ are the rotated masses of the $H$ doublet and $G$ doublet, respectively. This equation yields
\begin{eqnarray}
M_{H_b} - M_{H_c} & = & M_{G_b} - M_{G_c} = c_1 \left(\frac{1}{m_c} - \frac{1}{m_b} \right).
\end{eqnarray}
For the DHBs, their masses are expressed as
\begin{eqnarray}
M_{\mathcal{B}_+,cc} & = & \mathcal{M}_{\mathcal{B}_+} - \frac{2 \left( c_2 + c_3 \right) }{ m_c } \ , \nonumber\\
M_{\mathcal{B}_+,bc} & = & \mathcal{M}_{\mathcal{B}_+} - \left( c_2 + c_3 \right) \left( \frac{1}{m_c} + \frac{1}{m_b} \right) \ , \nonumber\\
M_{\mathcal{B}_+,bb} & = & \mathcal{M}_{\mathcal{B}_+} - \frac{2 \left( c_2 + c_3 \right) }{ m_b} \ ,
\label{HF violation 1}
\end{eqnarray}
which yield the mass relation
\begin{eqnarray}
M_{\mathcal{B}_+,bc} & = & \frac{1}{2}\left(M_{\mathcal{B}_+,cc} + M_{\mathcal{B}_+,bb}\right).
\label{eq:bc mass}
\end{eqnarray}
Similar relations hold for negative parity baryon $\mathcal{B}_-$. Explicitly
\begin{align}
M_{\mathcal{B}_-,cc} & =  \mathcal{M}_{\mathcal{B}_+} - \frac{2 \left( c_2 + c_3 \right) }{ m_c } \ , \notag\\
M_{\mathcal{B}_-,bc} & =  \mathcal{M}_{\mathcal{B}_+} - \left( c_2 + c_3 \right) \left( \frac{1}{m_c} + \frac{1}{m_b} \right) \ , \notag\\
M_{\mathcal{B}_-,bb} & =  \mathcal{M}_{\mathcal{B}_+} - \frac{2 \left( c_2 + c_3 \right) }{ m_b} \ .
\label{HF violation 2}
\end{align}
which yield
\begin{eqnarray}
M_{\mathcal{B}_-,bc} & = & \frac{1}{2}\left(M_{\mathcal{B}_-,cc} + M_{\mathcal{B}_-,bb}\right).
\end{eqnarray}
It should be stressed that the heavy-flavor symmetry leads to the degeneracy of two different heavy quark multiplets of $bc$ baryons, i.e., the heavy-quark singlet $\phi_{bc}$ type and the heavy-quark doublet $\psi_{bc}$ type, as pointed in Ref.~\cite{Ma:2015lba}. Here, above results in Eqs.~(\ref{HF violation 1}) and (\ref{HF violation 2}) show that the violating terms of the heavy-flavor symmetry at first order in Eq.~(\ref{eq:mQCor}) do not generate the mass difference between the $\phi_{bc}$-type baryons and $\psi_{bc}$-type baryons.

We would like to stress that the superflavor symmetry~\cite{Georgi:1990ak,Carone:1990pv} links the coupling constants in heavy meson sector to the present DHB sector~\cite{Hu:2005gf,Mehen:2017nrh}. Therefore, we can make predictions on the properties of DHBs using the present information of heavy-light mesons. After chiral symmetry breaking which can be achieved by a suitable choice of the potential of the light meson sector, $S$, and therefore $M$ field, acquire vacuum expectation value $\langle M \rangle = f_\pi$ with $f_\pi$ being the pion decay constant. From the Lagrangian \eqref{eq:EffecL}, the mass splitting of the chiral partners are
\begin{eqnarray}
\Delta M_q & = & m_{\mathcal{B}_{q-}} - m_{\mathcal{B}_{q+}} = m_{G_q} -m_{H_q} .
\end{eqnarray}

In the following, we shall perform numerical analyses.  For a realistic analysis, we phenomenologically include the effect of heavy spin violation as well as that of explicit breaking of the chiral symmetry from the current quark masses of $u$, $d$ and $s$ quarks. Let us explain how to include those effects in the following.

In the present analysis, we estimate $\Delta M_q$ using the masses of charmed mesons as follows.
In $u$-quark sector, we use the masses of $D^0$, $D^\ast(2007)^0$, $D_0^\ast(2400)^0$ and $D_1(2430)^0$ listed in Table~\ref{tab:meson masses} to obtain
\begin{eqnarray}
M_{H_{u}} & = & 1971.34 \pm 0.04 \,\mbox{MeV}\ , \nonumber\\
M_{G_{u}} & = & 2400 \pm 28 \,\mbox{MeV} \ ,
\end{eqnarray}
which we also use for $d$-quark sector.
\begin{table}[htbp]
\caption{Relevant masses of charm and bottom mesons
}\label{tab:meson masses}
\begin{tabular}{l|c}
\hline\hline
Particle & Masses (MeV) \\
\hline
$D^0$ & $1864.83 \pm 0.05$ \\
$D^{\ast}(2007)^{0}$ & $2006.85 \pm 0.05$ \\
$D_0^{\ast}(2400)^0$ & $2318 \pm 29$ \\
$D_1(2430)^0$ & $2427 \pm 26 \pm 25$ \\
$D_s^{\pm}$ & $1968.28 \pm 0.10$ \\
$D_s^{\ast\pm}$ & $2112.1\pm0.4$ \\
$D_{s0}^\ast(2317)^{\pm}$ & $2317.7 \pm 0.6$ \\
$D_{s1}(2460)^{\pm}$ & $2459.5 \pm 0.6$ \\
$B^0$ & $5279.63 \pm 0.15$ \\
$B^{\ast}$ & $5324.65 \pm 0.25$ \\
$B_s^{0}$ & $5366.89 \pm 0.19$ \\
$B_s^{\ast}$ & $5415.4^{+1.8}_{-1.5}$ \\
\hline\hline
\end{tabular}
\end{table}
For $s$-quark sector, from the masses of $D_{s}^\pm$, $D_s^{\ast\pm}$, $D_{s0}^\ast(2317)^\pm$ and $D_{s1}(2460)^\pm$ listed in Table~\ref{tab:meson masses}, we have
\begin{eqnarray}
M_{H_s} & = & 2076.2 \pm 0.3  \,\mbox{MeV} \ , \notag\\
M_{G_s} & = & 2424.1 \pm 0.5  \,\mbox{MeV} \ .
\end{eqnarray}
Using these values we estimate
\begin{equation}
\Delta M_q = \left\{
        \begin{array}{ll}
        429 \pm 28\,{\rm MeV} \ , & \hbox{for $q=u,d$;} \\
        347.9 \pm 0.6 \,{\rm MeV}\ . & \hbox{for $q=s$.}
        \end{array}
      \right.
\label{eq:MassDif}
\end{equation}

To estimate the masses of the DHBs, in addition to the mass relation from the Lagrangian~\eqref{eq:EffecL} and \eqref{eq:mQCor}, we take some inputs. For the doubly charmed baryons $\Xi_{cc}$, we take the central value of recent LHCb data $m_{\Xi_{cc}^{++}} = 3621.40$~\cite{Aaij:2017ueg}. For the mass of $\Omega_{cc}$, we estimate it from the relation
\begin{eqnarray}
m_{\Omega_{cc}} - m_{\Xi_{cc}} & \simeq & m_{H_s} - m_{H_q}
 = 104.9 \pm 0.3 \,\mbox{MeV} \ ,
\end{eqnarray}
which yields $m_{\Omega_{cc}} \simeq 3726.3\,$MeV with omitting the error bar. To estimate the DHBs including a pair of bottom quarks, we take  $m_{\Xi_{bb}} = 10150~$MeV as reference value, which is the average of the central values obtained in Refs.~\cite{Karliner:2014gca,Brown:2014ena}. The mass of $\Omega_{bb}$ is estimated as $m_{\Omega_{bb}} \simeq m_{\Xi_{bb}} + m_{\Omega_{cc}} - m_{\Xi_{cc}} \simeq  10155 $~MeV, which is in agreement with the calculation of Ref.~\cite{Brown:2014ena}. For the DHBs in the same heavy quark multiplet, even though they have the same mass in the heavy quark limit, their mass difference can be estimated by using the heavy quark-diquark symmetry relation~\cite{Brambilla:2005yk,Fleming:2005pd} as follows:
\begin{eqnarray}
m_{\Xi_{cc\,3/2}} - m_{\Xi_{cc\,1/2}} & = & \frac{3}{4} \left( m_{D^{\ast0}} - m_{D^0} \right)  \simeq 106.5 \,\mbox{MeV} \  , \nonumber\\
m_{\Omega_{cc\,3/2}} - m_{\Omega_{cc\,1/2}} & = & \frac{3}{4} \left( m_{D_s^{\ast0}} - m_{D_s^0} \right)  \simeq 107.9 \,\mbox{MeV} \ , \notag\\
m_{\Xi_{bb\,3/2}} - m_{\Xi_{bb\,1/2}} & = & \frac{3}{4} \left( m_{B^{\ast}} - m_{B^0} \right)  \simeq  33.8\,\mbox{MeV} \  , \nonumber\\
m_{\Omega_{bb\,3/2}} - m_{\Omega_{bb\,1/2}} & = & \frac{3}{4} \left( m_{B_s^{\ast}} - m_{B_s^0} \right)  \simeq  36.4\,\mbox{MeV} \ .
\label{eq:massdiffintra}
\end{eqnarray}
With the above discussions, we predict the masses of the DHBs in Table~\ref{tab:sumb}.

%%%%%%%%%%%%%%%%%%%%%%%%%%%%%%%%%%%%%%%%%%%%%%%%%%%%%%%%%%%%%%%%%%%%%%%%%%%%%%%%%%%%%%%%%%%%%%
\begin{table*}[htb]
%\begin{center}
\caption{\label{tab:sumb} Masses of the DHBs and the partial widths of one-pion
intermultiplet transitions and radiative intramultiplet transitions (in unit of MeV). For pion decays, the left values are for $g_S = 0.43$, while the right values are for $g_S = 1.63$.
For radiative decays, the left values are for $\beta ={} -2.9 \times 10^{-3}$\,MeV$^{-1}$ and the right ones for $\beta ={} - 0.21 \times 10^{-3}$\,MeV.
}
\begin{tabular}{c|c|c|c|c}
\hline
\hline
\qquad State \qquad\qquad & \ \ $J^P$ \ \ & \qquad Mass \qquad & \qquad Decay channel \qquad\qquad &
\qquad Partial width \qquad \\
\hline
$\Xi_{cc}$ & $\frac{1}{2}^+$ & $ 3621.4$~\cite{Aaij:2017ueg}  & --- & --- \\
\hline
$\Xi_{cc}^{\ast}$ & $\frac{1}{2}^-$ &  $4050$ & $\Xi_{cc}^{\ast ++} \to \Xi_{cc}^{++} + \pi^0$ & $100$, $92.6$ \\
& & & $ \Xi_{cc}^{\ast ++} \to \Xi_{cc}^{+} + \pi^+$ & $199$, $185$ \\
\hline
$\Xi_{cc}^\mu$ & $\frac{3}{2}^+$ & $3727.9$ & $\Xi_{cc}^{\mu ++} \to \Xi_{cc}^{++} + \gamma$ & $( 3.1\,,\,3.1) \times 10^{-3}$ \\
 & & & $\Xi_{cc}^{\mu +} \to \Xi_{cc}^{+} + \gamma$ & $( 15.6\,,\,4.8)\times 10^{-3}$
  \\
\hline
$\Xi_{cc}^{\prime \mu}$ & $\frac{3}{2}^-$ & $4157$ & $ \Xi_{cc}^{\prime\mu ++ }  \to \Xi_{cc}^{++ \mu} + \pi^0$ & $100$, $94.3$ \\
 & & & $  \Xi_{cc}^{\prime\mu ++ }  \to \Xi_{cc}^{+ \mu} + \pi^+$ & $199$, $188$  \\
~ & ~ & & $\Xi_{cc}^{\prime \mu ++} \to \Xi_{cc}^{\ast ++} + \gamma$ & $(3.1\,,\,3.2) \times 1-^{-3}$ \\
~ & ~ & & $\Xi_{cc}^{\prime\mu + } \to \Xi_{cc}^{\ast +} + \gamma$ & $(15.8\,,\,4.8)\times 10^{-3} $ \\
\hline
$\Omega_{cc}$ & $\frac{1}{2}^+$ & $3726.3$ & --- & --- \\
\hline
$\Omega_{cc}^{\ast}$ & $\frac{1}{2}^-$ & $4074.2$ & $ \Omega_{cc}^{\ast +} \to \Omega_{cc}^{+} + \pi^0 $ &  $( 2.9\,,\, 0.55) \times 10^{-3}$ \\
\hline
$\Omega_{cc}^{\mu}$ & $\frac{3}{2}^+$ &  $3834.2$ & $\Omega_{cc}^{\mu} \to \Omega_{cc} + \gamma$ & $(16.2\,,\, 5.0) \times 10^{-3} $ \\
\hline
$\Omega_{cc}^{\prime \mu}$ & $\frac{3}{2}^-$ & $4182.1$ & $ \Omega_{cc}^{\prime + \mu} \to \Omega_{cc}^{+ \mu} + \pi^0 $ & $( 2.9\,,\,0.55)\times 10^{-3}$ \\
~ & ~ & & $\Omega_{cc}^{\prime\mu} \to \Omega_{cc}^{\ast} + \gamma$ &  $(16.2\,,\,5.0 ) \times 10^{-3}$ \\
\hline
$\Xi_{bc}$ & $\frac{1}{2}^+$ & \  $ 6885.7 $  & --- & --- \\
\hline
$\Xi_{bc}^{\prime}$ & $\frac{1}{2}^+$ & $ 6885.7 $ & --- & --- \\
\hline
$\Xi_{bc}^{\ast}$ & $\frac{1}{2}^-$ &  $7315$ &$ \Xi_{bc}^{\prime\ast +} \to \Xi_{bc}^{\prime +} + \pi^0 $  & $104$, $114$ \\
&  & & $ \Xi_{bc}^{\ast +} \to \Xi_{bc}^{0} + \pi^+$ &$208$, $227$\\
\hline
$\Xi_{bc}^{\prime \ast}$ & $\frac{1}{2}^-$ & 7315 &  $ \Xi_{bc}^{\ast +} \to \Xi_{bc}^{+} + \pi^0 $ & $104$, $114$  \\
&  & & $ \Xi_{bc}^{\ast +} \to \Xi_{bc}^{0} + \pi^+$ & $208$, $227$   \\
\hline
$\Xi_{bc}^\mu$ & $\frac{3}{2}^+$ &  $6955.8$ & $\Xi_{bc}^{\mu} \to \Xi_{bc}^{} + \gamma$ &  $(2.2\,,\,0.29)\times 10^{-3}$ \\
 & & & $\Xi_{bc}^{\mu +} \to \Xi_{cc}^{+} + \gamma$ &  $(2.5\,,\,0.10)\times10^{-3}$ \\
\hline
$\Xi_{bc}^{\prime \mu}$ & $\frac{3}{2}^-$ &  $7384$& $\Xi_{bc}^{\prime + \mu} \to \Xi_{bc}^{+ \mu} + \pi^0$ &  $105$, $112$ \\
& & & $\Xi_{bc}^{\prime + \mu} \to \Xi_{bc}^{0 \mu} + \pi^+$ &  $208$, $224$ \\
~ & ~ & & $\Xi_{bc}^{\prime \mu +} \to \Xi_{bc}^{\ast + } + \gamma$ & $(2.4\,,\,0.096)\times10^{-3}$ \\
~ & ~ & & $\Xi_{bc}^{\prime \mu} \to \Xi_{bc}^{\ast} + \gamma$ & $(2.1\,,\,0.28)\times10^{-3}$ \\
\hline
$\Omega_{bc}$ & $\frac{1}{2}^+$ & $6940.7$ & --- & --- \\
\hline
$\Omega_{bc}^\prime$ & $\frac{1}{2}^+$ & $6940.7$ & --- & --- \\
\hline
$\Omega_{bc}^{\ast}$ & $\frac{1}{2}^-$ & $7288.6$ & $\Omega_{bc}^{\ast} \to \Omega_{bc} + \pi^0$ & $(3.0\,,\,0.67)\times 10^{-3}$ \\
\hline
 $\Omega_{bc}^{\prime\ast}$  & $\frac{1}{2}^-$  & $7288.6$  & $\Omega_{bc}^{\prime\ast} \to \Omega^{\prime}_{bc} + \pi^0$  &   $(3.0\,,\,0.67)\times 10^{-3}$  \\
\hline
$\Omega_{bc}^{\mu}$ & $\frac{3}{2}^+$ &  $7012.8$ & $\Omega_{bc}^{\mu} \to \Omega_{bc}^{} + \gamma$ &  $(2.4\,,\,0.32)\times10^{-3}$ \\
\hline
$\Omega_{bc}^{\prime \mu}$ & $\frac{3}{2}^-$ &  $7360.7$ & $ \Omega_{bc}^{\prime \mu} \to \Omega_{bc}^{ \mu} + \pi^0$ & $(3.1\,,\,0.67)\times 10^{-3}$ \\
~ & ~ & &  $\Omega_{bc}^{\prime\mu} \to \Omega_{bc}^{\ast} + \gamma$ & $(2.4\,,\,0.32)\times10^{-3}$ \\
\hline
$\Xi_{bb}$ & $\frac{1}{2}^+$ & ~$ 10,150 $~\cite{Karliner:2014gca,Brown:2014ena} & --- & --- \\
\hline
$\Xi_{bb}^{\ast}$ & $\frac{1}{2}^-$ & $10,579$ & $ \Xi_{bb}^{\ast 0} \to \Xi_{bb}^{0} + \pi^0$ & $106$, $122$ \\
 & & & $ \Xi_{bb}^{\ast 0} \to \Xi_{bb}^{-} + \pi^+$ &  $211$, $243$ \\
\hline
$\Xi_{bb}^\mu$ & $\frac{3}{2}^+$ & $10,184$ & $\Xi_{bb}^{\mu} \to \Xi_{bb}^{} + \gamma$ &  $(0.56\,,\,0.011)\times10^{-3}$ \\
 & & & $\Xi_{bb}^{\mu -} \to \Xi_{bb}^{-} + \gamma$ & $(0.17\,,\,0.0021)\times10^{-3}$  \\
\hline
$\Xi_{bb}^{\prime \mu}$ & $\frac{3}{2}^-$ &  $10,613$& $ \Xi_{bb}^{\prime 0} \to \Xi_{bb}^{0 \mu} + \pi^0$ & $106$, $122$ \\
&  & & $ \Xi_{bb}^{\prime 0} \to \Xi_{bb}^{- \mu} + \pi^+$ &  $211$, $243$ \\
~ & & ~ & $\Xi_{bb}^{\prime \mu} \to \Xi_{bb}^{\ast} + \gamma$ &  $(0.56\,,\,0.011)\times10^{-3}$ \\
~ & & ~ & $\Xi_{bb}^{\prime \mu - } \to \Xi_{bb}^{\ast -} + \gamma$ &  $(0.17\,,\,0.0021)\times10^{-3}$ \\
\hline
$\Omega_{bb}$ & $\frac{1}{2}^+$ &  $10,155$ & --- & --- \\
\hline
$\Omega_{bb}^{\ast}$ & $\frac{1}{2}^-$ & $10,503$ & $ \Omega_{bb}^{\ast -} \to \Omega_{bb}^{-} + \pi^0 $ & $(3.1\,,\,0.72)\times 10^{-3}$ \\
\hline
$\Omega_{bb}^{\mu}$ & $\frac{3}{2}^+$ & $10,191$ & $\Omega_{bb}^{\mu} \to \Omega_{bb}^{} + \gamma$ &  $(0.098\,,\,0.0012)\times10^{-3}$\\
\hline
$\Omega_{bb}^{\prime \mu}$ & $\frac{3}{2}^-$ & $10,539$& $ \Omega_{bb}^{\prime - \mu} \to \Omega_{bb}^{- \mu} +\pi^0 $ & $(3.1\,,\,0.72)\times 10^{-3}$ \\
~ & & ~ & $\Omega_{bb}^{\prime\mu} \to \Omega_{bb}^{\ast} + \gamma$ & $(0.098\,,\,0.0012)\times10^{-3}$ \\
\hline
\hline
\end{tabular}
%\end{center}
\end{table*}
%%%%%%%%%%%%%%%%%%%%%%%%%%%%%%%%%%%%%%%%%%%%%%%%%%%%%%%%%%%%%%%%%%%%%%%%%

We next turn to the strong decays of the DHBs. For this purpose, we first estimate the coupling constant $g_\pi$ which measures the intermultiplet transitions of the chiral partners. From the mass splitting for $q = u,d$ in Eq.~\eqref{eq:MassDif}, one finds that $g_\pi$ can be estimated by using the generalized Goldberger-Treiman relation
\begin{eqnarray}
g_\pi & = & \frac{\Delta M_q}{f_\pi} = 4.6 \pm 0.3 \ ,
\label{gpi val}
\end{eqnarray}
where we use $f_\pi = 92.42 \pm 0.26$\,MeV. Then, by using this value and considering the isospin relation, one can calculate the full width of $D_0^\ast \to D \pi$ as
\begin{eqnarray}
\Gamma(D_0^\ast \to D \pi) & = & \frac{3}{2}\left(2g_\pi^2 m_{D_0^\ast} m_{D}\right)\frac{1}{8\pi m_{D_0^\ast}^2}|\vec{p}_\pi| \nonumber\\
& = & 793 \pm 103 \, \mbox{MeV} \ ,
\end{eqnarray}
where $|\vec{p}_\pi|$ is the three momentum of the decay products. The above result is about three times of the observed widths $267\pm 40\,$MeV ($D_0^{\ast0}$) and $230\pm17$\,MeV ($D_0^{\ast\pm}$)~\cite{Olive:2016xmw}, so that, if one only used the $g_\pi$ term in Lagrangian~\eqref{eq:EffecL} to calculate the one-pion intermutiplet transition, the predicted widths of DHBs would be too large. Considering this defect, we include the $g_S$ term in the Lagrangian~\eqref{eq:EffecL}. Therefore, the full width of $D_0^\ast \to D \pi$ is written as
\begin{eqnarray}
\Gamma(D_0^\ast \to D \pi) & = &
\frac{3}{8\pi} \left( g_\pi - \frac{g_S E_\pi}{ f_\pi } \right)^2 \frac{m_D}{ m_{D_0^\ast} } \vert \vec{p}_\pi \vert \ ,
\end{eqnarray}
where $E_\pi$ is the energy of the outgoing pion. By using the experimental value of the total width of $D_0^\ast(2400)^0$, $267\pm 40$\,MeV~\cite{Olive:2016xmw},
 one obtains two solutions as
 \begin{equation}
g_S = 0.43 \pm 0.08 \ , \quad 1.63 \pm 0.14 \ ,
\label{gS val}
\end{equation}
which are common in the calculation of the intermultuplet decays of DHBs.

In terms of the mass difference $\Delta M_{B;q}$, the intermultiplet one-pion transitions of the DHBs in the isospin symmetry limit can be studied. The relevant partial widths are expressed as
%\begin{widetext}
\begin{eqnarray}
\Gamma \left( \Xi_{QQ^\prime}^{\ast} \to \Xi_{QQ^\prime} \pi \right) & \simeq & \Gamma \left(
\Xi_{QQ^\prime}^{\prime \mu } \to \Xi_{QQ^\prime}^{\mu } \pi \right) \nonumber\\
& \simeq & \Gamma \left( \Xi_{QQ^\prime}^{\prime\ast } \to \Xi_{QQ^\prime}^{\prime} \pi \right) \nonumber\\
& = & \frac{3}{8\pi} \left( g_\pi - \frac{g_S E_\pi}{ f_\pi } \right)^2 \frac{m_{\Xi_{QQ}} }{ m_{\Xi_{QQ'}^\ast} } \vert \vec{p}_\pi \vert \ , \nonumber\\
\Gamma \left( \Omega_{QQ^\prime}^{\ast} \to \Omega_{QQ^\prime} \pi \right) & \simeq & \Gamma \left(
\Omega_{QQ^\prime}^{\prime \mu } \to \Omega_{QQ^\prime}^{\mu } \pi \right) \nonumber\\
& \simeq & \Gamma \left( \Omega_{QQ^\prime}^{\prime\ast } \to \Omega_{QQ^\prime}^{\prime} \pi \right) \nonumber\\
& = & \frac{1}{8\pi} \left( g_\pi - \frac{g_S E_\pi}{ f_\pi } \right)^2 \frac{m_{\Omega_{QQ}} }{ m_{\Omega_{QQ'}^\ast} } \vert \vec{p}_\pi \vert \Delta_{\pi^0\eta}^2 ,
\nonumber\\
\label{eq:1pionXi}
\end{eqnarray}
%\end{widetext}
where $\vert p_\pi \vert $ and $E_\pi$ are the three-momentum and energy of $\pi$ in the rest frame of the decaying DHB and $\Delta_{\pi^0\eta}$ is the magnitude of the $\eta$-$\pi^0$ mixing. Here, we take  $\Delta_{\pi^0\eta} = {} -5.32 \times 10^{-3}$ estimated in Ref.~\cite{Harada:2003kt} based on the two-mixing angle scheme (see, e.g., Ref.~\cite{Harada:1995sj} and references therein). The channels including charged pions can be obtained by using the isospin relation. Our results of the intermultiplet one-pion transition are summarized in Table~\ref{tab:sumb}.

Since the mass difference between the DHB in a heavy quark multiplet with spin-$3/2$ and that with spin-$1/2$ is less than the pion mass, unlike the heavy-light meson including a charm quark, the intramultiplet hadronic transition is forbidden. Therefore, the dominant intramultiplet transition is the electromagnetic transition. In our present framework, we can write the Lagrangian for the magnetic decays of heavy hadrons as
\begin{eqnarray}
{\mathcal L} & = & \frac{e \beta}{2} \, \mbox{Tr} \Big[ \bar{\Psi}_L \sigma_{\rm light}^{\mu\nu} F_{\mu\nu} Q_{\rm light} \Psi_L  \nonumber\\
& & \qquad\quad{} +  \bar{\Psi}_R \sigma_{\rm light}^{\mu\nu} F_{\mu\nu} Q_{\rm light} \Psi_R
\Big] \nonumber\\
& &{} + e \, \mbox{Tr} \left[ \sigma_{\rm heavy}^{\mu\nu} F_{\mu\nu} Q_{\rm heavy} \, \left( {\mathcal H}_L \bar{\mathcal H}_L  +  {\mathcal H}_R \bar{\mathcal H}_R \right) \right]  \nonumber\\
& &{}
+ e \, \mbox{Tr} \left[ \sigma_{\rm heavy}^{\mu\nu} F_{\mu\nu} Q_{\rm heavy}  \left( \bar{\mathcal B}_L {\mathcal B}_L  +  \bar{\mathcal B}_R {\mathcal B}_R \right) \right]  \nonumber\\
& &{}
+ e \, \mbox{Tr} \bigg[ \bar{\mathcal B}_L \, \sigma_{\rm heavy}^{\mu\nu} F_{\mu\nu} Q_{\rm heavy}  \, {\mathcal B}_L  \nonumber\\
& & \qquad\qquad {} +  \bar{\mathcal B}_R  \, \sigma_{\rm heavy}^{\mu\nu} F_{\mu\nu} Q_{\rm heavy}  \, {\mathcal B}_R  \bigg]  \ ,
\end{eqnarray}
where $\beta$ is the parameter introduced to account for the magnetic moment of the light quark in the heavy hadron, $Q_{\rm light} = {\rm diag}(2/3,-1/3,-1/3)$ is the charge matrix of the light quark and $Q_{\rm heavy} = {\rm diag}(2/3/m_c,-1/3/m_b)$ denotes the charges of the heavy quark in the heavy hadron. $F_{\mu\nu}$ is the field strength of the photon field.  The subscript ``light'' for $\sigma^{\mu\nu} = \frac{i}{2} \left[ \gamma^\mu , \gamma^\nu \right]$ implies that it acts on the spinor indices for light quarks, and the subscript ``heavy'' does for heavy quarks. Using this Lagrangian, one can obtain the following simple rates for the intramultiplet transitions
\begin{eqnarray}
& & \Gamma\left(1^{\pm} \to 0^{\pm} \gamma\right) = \frac{1}{3}\alpha_{\rm em}
\frac{m_{{\mathcal H}^{(\ast)}_{\bar{Q} q}}}{m_{{\mathcal H}^{(\prime)}_{\bar{Q} q}}}
\left( e_q \, \beta - \frac{e_Q}{m_Q}  \right)^2
\left|\vec{P}_\gamma\right|^3,\nonumber\\
\label{meson photon}
\\
& & \Gamma\left(\frac{3}{2}^{\pm} \to \frac{1}{2}^{\pm} \gamma\right) = \frac{4}{9}\alpha_{\rm em}
\frac{m_{{\mathcal B}^{(\ast)}_{QQ^\prime q}}}{m_{{\mathcal B}^{(\prime)\mu}_{QQ^\prime q}}}
\nonumber\\
& &\qquad\qquad\qquad\qquad {} \times
\left( e_q \, \beta + \frac{e_Q}{m_Q} + \frac{e_{Q^\prime}}{m_{Q^\prime}} \right)^2
\left|\vec{P}_\gamma\right|^3,
\label{baryon photon}
\end{eqnarray}
where $|\vec{P}_\gamma|$ is the three momentum of photon. Instead of determine $\beta$ trough relation $\beta = 1/m_q$ with $m_q$ being the light constituent quark mass, we determine the value of $\beta$ from Eq.~(\ref{meson photon}) using the experimental data $\Gamma(D^{\ast\pm} \to D^{\pm} + \gamma) = 1.3\,$KeV with $m_c = 1.28$\,GeV as
\begin{equation}
\beta = \left( - 2.9 \, ,\, -0.21 \right) \times 10^{-3} \, \mbox{MeV}^{-1}\ ,
\end{equation}
where the former value is similar to that determined by using constituent light quark mass. Then, using this value in Eq.~(\ref{baryon photon}), we calculate the radiative decays of doubly heavy baryons which are listed in Table~\ref{tab:sumb}.

In this work, considering the recent LHCb observation, we studied the spectrum of doubly heavy baryons based on the chiral partner structure model. The masses of the spin-parity $(1/2)^{\pm}, (3/2)^{\pm}$ DHBs are estimated, the intermultiplet one-pion transitions of the DHBs are calculated and the electromagnetic intramultiplet transitions are calculated. Similar to our previous work, the splitting of the masses of chiral partners are estimated to be $430$\, MeV for the DHBs including up or down quarks which $350~$MeV for DHBs including strange quark. Using the extended Lagrangian in this work, we calculated the intermultiplet one-pion transitions and found that the total widths of the DHBs with negative parity are revised to $\sim 300$\,MeV for the DHBs including up or down quarks but those for the DHBs including a strange quark are revised to $\lesssim 3\,$KeV. In addition, we also calculated the intramultiplet electromagnetic transition of the DHBs, since unlike the charmed mesons, the mass difference between the heavy quark partner is smaller than the pion mass. The present results can be used as a hints for the future DHB search. It will be interesting to study the modification of masses of chiral partners of DHBs similarly to Refs.~ \cite{Suenaga:2014sga,Harada:2016uca,Suenaga:2017deu}.

%%%%%%%%%%%%%%%%%%%%%%%%%%%%%%%%%%%%%%%%%%%%%%%%%%%%%%%%%%%%%%%%%%%%%%%%%
\subsection*{Acknowlegments}

The work of M. H. was supported in part by the JSPS
Grant-in-Aid for Scientific Research (C) No. 16K05345.
Y.~L. M. was supported in part by National Science
Foundation of China (NSFC) under Grant No. 11475071, 11547308 and the Seeds Funding of Jilin
University.

%%%%%%%%%%%%%%%%%%%%%%%%%%%%%%%%%%%%%%%%

\end{document}